\documentclass[11pt,a4paper]{article}
\usepackage[left=1in, right=1in, top=1in, bottom=1in]{geometry}

\usepackage{authblk}
\usepackage{enumitem}
\usepackage{tabularx}
\usepackage[dvipsnames]{xcolor}
\usepackage{hyperref}
\usepackage{multirow}

\usepackage{amsmath,amssymb,amsfonts,mathtools}
\usepackage[mathscr]{euscript}
\usepackage{graphicx}
\graphicspath{ {./images/} }

\definecolor{red}{cmyk}{0,1,0.77,0.3}
\definecolor{yellow}{cmyk}{0,0.05,0.89,0.03}
\definecolor{green}{cmyk}{0.65,0,0.92,0.15}
\definecolor{blue}{cmyk}{0.9,0.3,0.2,0.07}
\definecolor{grey}{cmyk}{0,0,0,0.5}

\newcommand{\GG}{\textbf{\textcolor{green} {G}}}
\newcommand{\BB}{\textbf{\textcolor{blue} {B}}}
\newcommand{\RR}{\textbf{\textcolor{red} {R}}}
\newcommand{\gG}{\textbf{\textcolor{grey} {g}}}
\newcommand{\gB}{\textbf{\textcolor{grey} {b}}}
\newcommand{\gR}{\textbf{\textcolor{grey} {r}}}

\newcommand*{\email}[1]{\href{mailto:#1}{\url{#1}} }

\begin{document}

\title{Sliced, not Splitted: a Better Alternative to Many-Worlds?}
\date{\vspace{-5ex}}

\author[1]{\small
Christopher Thron
and Braeden Welsch 
}
\affil[1]
{\small Department of Science and Mathematics, Texas A\&M University-Central Texas\\
\email{thron@tamuct.edu}}

\maketitle

\noindent{\bf Abstract:}  
The many-worlds interpretation (MWI) of quantum mechanics is currently experiencing a popular resurgence, propelled by such prominent and articulate physicists as  Sean Carroll, David Deutsch, Max Tegmark, and Lev Vaidman. The consequences of MWI are mind-boggling: the spacetime universe of our experience is only one branch of an unimaginably fast-multiplying plethora of alternative universes held incommunicado.    
In this paper, we propose that the mass of Medusa's hair served up by MWI is  due to a failure to embed the spacetime universe in the right space: spacetime is a slice of bread, not a splitting strand of pasta. By way of motivation, we first give a very simple presentation of Bell's inequality by comparing it to a ``quantum game show'', followed by a simple description of Aspect's 1985 experiment involving entangled photons which confirms the inequality. We interpret the paradoxical correlation between measurements as resulting from a process outside of spacetime that produces both the original entanglement and the measurements. This is followed by a brief presentation of MWI, and then by a pictorial comparison of the proposed process model and MWI.  The final section lists a number of potential consequences of the model related to causality, determinism, free will, and consciousness, and points to further references that give a more in-depth and rigorous presentation of the process dimension model.

The entire article is non-technical and requires no mathematical background other than high school mathematics and an understanding of basic concepts in probability. The physics involved in Aspect's experiment is also explained. 

\noindent{\bf Keywords:} quantum mechanics; interpretation; Bell's inequality; Aspect experiment; entanglement; correlation; causality; process; spacetime; universe; many-worlds; free will.\\


\maketitle

\section{A quantum game show}\label{sec:quantumGame}

\begin{figure}[h]
    \begin{center}
        \includegraphics[scale=0.5]{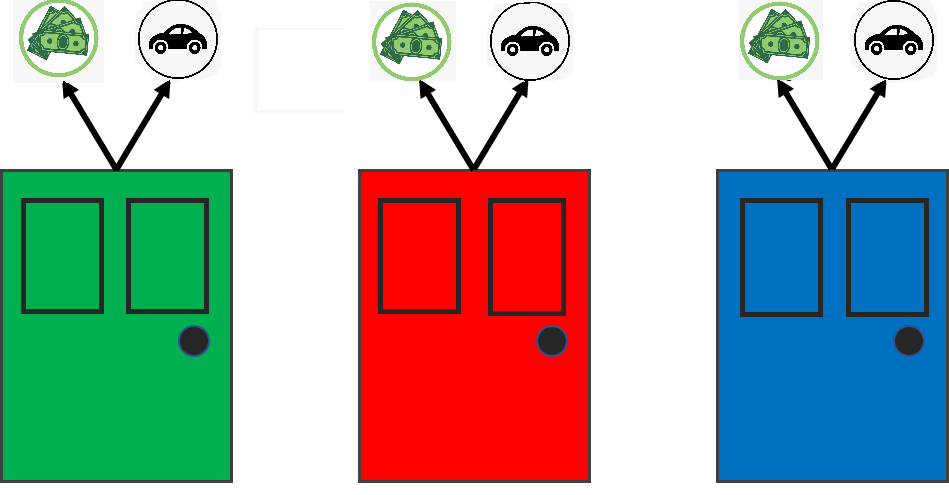}
    \end{center}
    \caption{The doors, in order from left to right, are green (\GG), red (\RR), and blue (\BB). The arrows point to the two possible prizes (cash or car), one of which is behind each door.}\label{fig:doors}
\end{figure}

\quad~ The following analogy will elucidate a fundamental difference between classical and quantum mechanics. The example involves a game show in which a single participant chooses two of three doors, each concealing one of two prizes (cash or car). Let us label our three doors as green (\GG), red (\RR), and blue (\BB). Figure~\ref{fig:doors} shows the setup.

The goal of this ``Quantum Game Show" is to choose two doors with the \emph{same} prize (similar to Wheel of Fortune's ``half-car" prize, where you only win the car if you get both halves). Dedicated fans of the show who've watched thousands of episodes have calculated the probability of winning by choosing adjacent doors is $85\%$, For example, since the green door \GG~ is next to the red door \RR, it follows that that the probability that the green and red prizes are the same is $0.85$. For short, we'll write this as $P(\GG = \RR)=0.85$  Similarly, we also have
$P(\RR=\BB)=0.85$.

Now comes the big question. Knowing this information (and only this information), can we estimate  the probability of winning by choosing the green and blue doors (i.e. $P(\GG=\BB)$)?  

Let's consider the possibilities. A little thought reveals that it must be true that at least two doors have the same prize at all times, and possibly all three doors have the same prize. 
To represent the different possibilities (and to brighten up the text), let's introduce some colorful notation. 
We'll  indicate which door has a different prize from the other two by greying out its color and using lowercase. For example, \GG\RR\gB~ indicates the case where the green and red doors have the same prize, and the blue door has a different prize. Alternatively, \GG\RR\BB~ is the case where all three doors have the same prize. 
Table~\ref{tab:RGB} summarizes all possible arrangements.

\begin{table}[h!]
    \centering
\begin{tabular}{cc|cc}
     & \multicolumn{3}{r}{ \footnotesize{\GG~\& \RR~outcomes~~~~~}}\\
     & &
     $~~\GG = \RR~~$ & $~~\GG \neq \RR~~$    \\ \cline{2-4}
     \multirow{2}{4em}{ \footnotesize{\RR\&\BB\\~outcomes}} & 
     $\RR = \BB$  & \GG\RR\BB & \gG\RR\BB \\
  & $\RR \neq \BB$
  & \GG\RR\gB & \GG \gR\BB \\
\end{tabular}
\    \caption{Possible arrangements of matching prizes behind the three doors. For example,  \gG\RR\BB~ represents the case when the red and blue doors have the same prize  while the green door is different.}
    \label{tab:RGB}
\end{table}


The four entries in the table represent all possible prize arrangements, and exactly one of these arrangements must be realized each time the game is played. It follows that the sum of the probabilities of the four outcomes must be 1:

\begin{equation}\label{eq:1}
    P(\GG\RR\BB)+P(\GG\RR\gB)+P(\gG\RR\BB)
        + P(\GG \gR\BB)
      =1
\end{equation}

The first column of Table~\ref{tab:RGB} lists the two possible ways that the green prize can be the same as the red prize. As mentioned above, game fans have determined the probability of this happening is 0.85. We may then add the probabilities and obtain:

\begin{equation}\label{eq:2}
 P(\GG\RR\BB)+P(\GG\RR\gB) = P(\GG =\RR) = 0.85 
 \end{equation}

Also, the first row of Table~\ref{tab:RGB} lists the two possible ways that the red prize can be the same as the blue prize. We know already this happens with probability 0.85. So by similar reasoning we have

\begin{equation}\label{eq:3}
 P(\GG\RR\BB)+P(\gG\RR\BB) = P(\RR=\BB) = 0.85. 
 \end{equation}

Now for some algebra. 
Taking \eqref{eq:2} + \eqref{eq:3} $-$ \eqref{eq:1} and cancelling terms gives us:
\begin{equation*}
P(\GG\RR\BB) - P(\GG \gR\BB)= 0.7, 
\end{equation*}
and adding $2P(\GG \gR\BB)$ to both sides of the equation gives:
\begin{equation}\label{eq:Bell1}
P(\GG\RR\BB) + P(\GG \gR\BB)= 0.7 + 2P(\GG \gR\BB). 
\end{equation}
The left-hand side of \eqref{eq:Bell1} represents all possible arrangements where \GG~ is equal to \BB~ (i.e. $\GG=\BB$); while the right hand is 0.7 plus a non-negative number, which must be greater than or equal to 0.7. This leads us to our final result: 

\begin{equation}\label{eq:Bell2}
    P(\GG=\BB) \geq 0.7.
\end{equation}
In conclusion, the probability of having the same prize behind the green and blue doors must be at least $70\%$. So although we can't find the exact probability, at least we can get a lower estimate.

If we'd worked out the formula without putting in specific numbers, we'd have found that in general:
\begin{equation}\label{eq:Bell3}
    P(\GG=\BB) = P(\GG=\RR) + P(\RR=\BB) - 1 + 2P(\GG\gR\BB),
\end{equation}
or 
\begin{equation}\label{eq:Bell4}
    P(\GG=\BB) \ge P(\GG=\RR) + P(\RR=\BB) - 1.
\end{equation}

Equation~\eqref{eq:Bell4} is one way of stating \emph{Bell's inequality}. 

 There are no tricks in what we've done so far--it's all up-and-up mathematics, straight out of any probability textbook. But what's actually measured in the real-life quantum experiments doesn't agree, as we shall see shortly.

For an alternative presentation of Bell's inequality (without the game show, and with more physics) see 
 Lorenzo Maccone's article  \cite{maccone2013simple}. 
 
\section{Aspect's Experiment}

A real-life ``Quantum Game Show" experiment was conducted by French physicist Alain Aspect during the early 1980s \cite{aspect1982experimental}. 
In order to understand the experiment, we first need some  background information about physics. The following sections introduce two key physical phenomena that play important roles in Aspect's experiment: polarization and photons.

\subsection*{Polarization of light waves}
In physics, there are two basic wave types: longitudinal and transverse. Both types of waves have a direction of travel and direction of oscillation (the ``back-and-forth'' motion of the wave). The relation between these directions determines which type of wave it is. Longitudinal waves travel \emph{parallel} to the direction of oscillation, whereas transverse waves travel \emph{perpendicular} to the direction of oscillation. Traffic jams and sound are common examples of longitudinal waves, while light and ripples on the surface of water are examples of transverse waves. 

\begin{figure}[h]
    \begin{center}
        \includegraphics[scale=0.5]{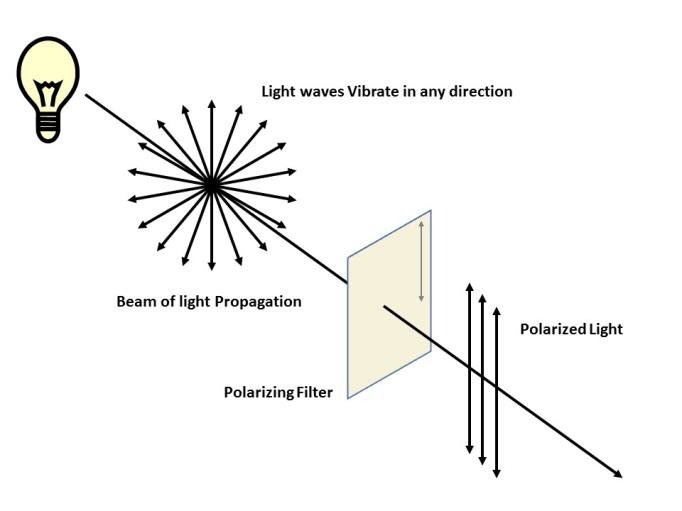}
    \end{center}
    \caption{Light source producing unpolarized light which passes through a polarizer to produce a polarized light wave. Web Reference:
    \url{https://www.researchgate.net/publication/328100369_Polarised_infrared_light_enables_enhancement_of_histo-morphological_diagnosis_of_prostate_cancer}}
    \label{fig:unpolarized}
\end{figure}

Polarization is a property of transverse waves in which the direction of oscillation is consistent. For example, water waves are polarized because the oscillation is always in the up-and-down direction, and never from side to side. This is not true of light: for example the Sun (and other common light sources) emits \emph{unpolarized} light waves, or \emph{electromagnetic waves}, which we see in Figure \ref{fig:unpolarized}. In this case, the direction of oscillation can be random, as long as  it's always perpendicular to the direction of travel. Unpolarized light waves can become polarized by passing through a \emph{polarizer} which fixes the oscillation in a consistent direction. 

\begin{figure}[h]
    \begin{center}
        \includegraphics[scale=0.95]{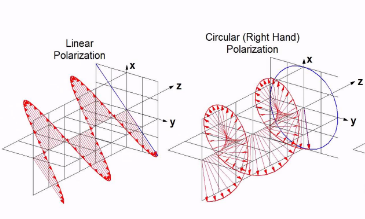}
    \end{center}
    \caption{Linear and circular polarization of electromagnetic waves. The red arrow in each diagram shows the direction of the electric field in the electromagnetic wave (light wave). Taken from \url{https://www.emedicalprep.com/study-material/physics/wave-optics/polarization/}}
    \label{fig:waves}
\end{figure}

There is more than one way that a transverse wave can be polarized. Linear polarization is the simplest type of polarization. Suppose you are holding one end of a long rope and begin to move your hand directly up and down. The rope's consequential oscillating motion corresponds to a linearly polarized transverse wave. A mathematician would recognize the produced wave as a sinusoidal wave. If you were to move your hand continuously around in a circle, the rope's oscillation would then represent a circularly polarized transverse wave. Figure \ref{fig:waves} demonstrates these two distinct types of  polarized transverse waves (it's  also possible to have elliptical polarization, but that's more complicated than we need to consider right now).

Light beams should actually not be considered as a single wave, but as a collection of many individual waves travelling in the same direction. Each individual wave carries a small amount of energy: the amount of energy is determined by frequency of the wave (the frequency of the wave also determines its color: for example, red light has a lower frequency (and less energy) than blue light). Each one of these individual waves is called a \emph{photon}. Photons are often thought of as ``particles'' of light, and in some respects they behave like  discrete objects. However, this conception is inaccurate and misleading, so we will avoid referring to photons in this way. 

Each individual photon has a definite polarization, but a light beam may consist of many photons with different polarizations that happen to be traveling in the same direction.
If you send a single photon through a polarizer, it either makes it through or it doesn't. If the photon makes it through, then the outgoing photon has a polarization that is parallel to the polarizer's orientation. If the photon doesn't make it through, then the photon is  reflected by the polarizer, and its outgoing polarization is perpendicular to the polarizer's orientation. 

If a light beam consists of circularly polarized photons, then regardless of polarizer's orientation half of the photons will be measured as parallel  and half will be measured as perpendicular. A key point here is that the measurement of a circularly polarized photon necessarily changes the polarization of the photon. Unlike ordinary `classical' physics, in quantum physics it is impossible to make a measurement without changing the thing you're measuring: this is one of the biggest differences between classical and quantum physics.  

\subsection*{The Experiment}

\begin{figure}[h]
    \begin{center}
        \includegraphics[scale=0.5]{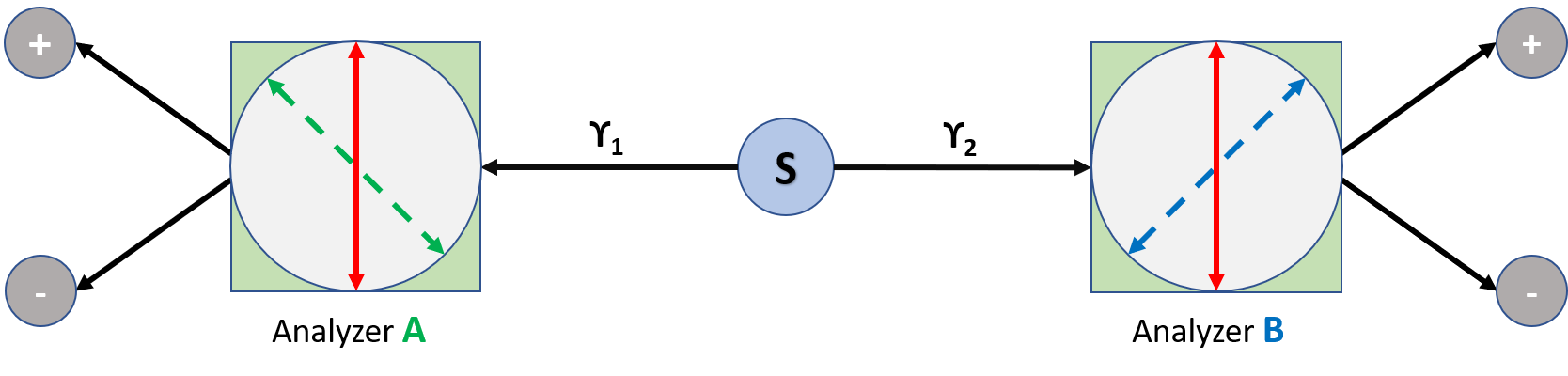}
    \end{center}
    \caption{Schematic representation of Aspect's experiment, as described in \cite{aspect1982experimental}. The red line indicates the initial orientation of $A$ and $B$. The green dashed line indicates $A$ is oriented in the counter-clockwise direction while the blue dashed line indicates $B$ is oriented in the clockwise direction. The angle between $A$ and $B$ is used to calculate the probability of the outcome. }
    \label{fig:setup}
\end{figure}

We now have the physical background needed to understand Aspect's experiment, which is schematically represented in Figure~\ref{fig:setup}. At the center of the experiment lies a source $S$ which prepares two 'identical' circularly-polarized photons $\gamma_1$ and $\gamma_2$, travelling in opposite directions. These photons are `identical' in the sense that their polarizations are the same according to any observer who measures them (in physics terminology, the two photons are said to be \emph{entangled}). 

For this experiment, the observers are linear polarizers placed on opposite sides of the source which measure the two photons' polarizations as they pass through. Let $A$ be the polarizer that measures $\gamma_1$ and $B$ be the polarizer that measures $\gamma_2$. There are two possible measurements $A$ and $B$ can give for their respective photons: parallel $(+)$ or perpendicular $(-)$. For example, if $A$ reads $(+)$, then $\gamma_1$'s polarization is parallel relative to the $A$'s orientation. 

Initially, $A$ and $B$ are oriented in the same direction and thus yield the same measurement, since $\gamma_1$ and $\gamma_2$ are entangled. In this case, the angular difference between $A$ and $B$ (denoted by $\theta$) is zero. We thus have $P_{same}(\theta)=1$ for $\theta=0$.

In general, given the angle between $A$'s and $B$'s orientation is $\theta$, quantum theory predicts that the probability that $A$ and $B$ have the same measurement is $P_{same}(\theta)=\cos^2{\theta}$.Since the photons either have the same or different polarizations, it follows that $P_{diff}(\theta)=1 - \cos^2{\theta} = \sin^2{\theta}$.  For a mathematical derivation of these results, see \cite{Aspectbrief}.

For the purposes of this experiment, we use three different orientations for $A$ and $B$ (we'll be referring to Figure \ref{fig:setup} above.)
\begin{enumerate}[label=(\arabic*)]
    \item Polarizer $A$ is rotated to align with green arrow while $B$ remains on the red arrow, where the angle between the two is $\theta=22.5^{\circ}$.
    
    \item Polarizer $A$ remains on the red arrow while $B$ is rotated to line up with  the blue arrow to give the angle $\theta=22.5^{\circ}$.
    
    \item Polarizers $A$ and $B$ are aligned with the green and blue arrows respectively, so that $\theta = 45^\circ$.
\end{enumerate}
In cases (1) and (2), we can easily see that the probability that $A$ and $B$ have the same measurement is $P_{same}(\theta=22.5^{\circ})=\cos^2{(22.5^{\circ})}=.85$, rounded to the nearest hundredth. Using the quantum game show probabilities 
derived in Section \ref{sec:quantumGame} 
we obtain the following result: 
\begin{align*}
\textrm{(According to game show)}\qquad
    P_{same}(45^\circ)  \geq .85+.85-1  =  0.7.
\end{align*}
Yet, when we use quantum mechanics to predict the angle of orientation between $A$ and $B$, we get:
\begin{align*}
\textrm{(According to quantum theory)}\qquad 
    P_{same}(45^\circ) =\cos^2(45^\circ)=0.5
\end{align*}

So who's right: quantum mechanics, or our intuition?  Both can't be right!  What Aspect showed is that quantum mechanics wins, and intuition fails. The probability turns out to be only $0.5$.

\subsection*{The bottom line}\label{sec:hidden}

What's going on here? It seems that we're missing something.
Whenever something happens that we don't expect, that means that we've assumed something which is not actually the case.  But what have we assumed?  It seems that in the game-show example the only assumption we've made is that the laws of probability are valid. But in fact, we've also assumed something else, namely that the game-show host is unaware of the choice we are going to make, so the prizes are set regardless of what doors (measurements) we choose.  

The violation of Bell's inequality demonstrates a similar hidden assumption in our analysis of Aspect's experiment.  Table~\ref{tab:RGB} assumes  that there are four definite, distinct possibilities for the two prepared photons.  In other words, the photons have ``already decided'' what they are going to do before the experimenter decides which measurement to make. By doing so, we assume the state they are in doesn't depend on which measurements are actually made. Mathematically, we describe this by saying that the photon states prior to measurement are \emph{independent} of which measurements are subsequently made. 

Based on our experience, this assumption is `obviously' true. How could a measurement of a photon reach back into the past and change how the photon was created? But let us recall Sherlock Holmes' famous dictum: ``When you have eliminated the impossible, whatever remains, however improbable, must be the truth.'' \cite{doyle2014}

\section{The process space proposal}
In the previous section, we suggest that perhaps we have to take another look at how we envision reality. One way to develop a new perspective is to look for possible analogies with situations that we're more familiar with. This is exactly what Einstein did when he viewed gravitation in terms of the ``curving'' or ``bending''  of spacetime, which can be compared to the curving of a rubber sheet in three-dimensional space. 

In this section we'll attempt to explain the ``quantum game show'' by taking the same approach. First we'll
describe a commonplace situation, and then proceed to develop possible connections between this well-understood occurrence and the puzzling situation we see with quantum reality. In the next section,  we'll  compare this  approach  with another popular explanation of the ``quantum game show'', namely the so-called the ``many-worlds interpretation''. See which you think is better!     

\subsection*{A pebble in a pond}
\begin{figure}[h]
    \begin{center}
        \includegraphics[scale=0.5]{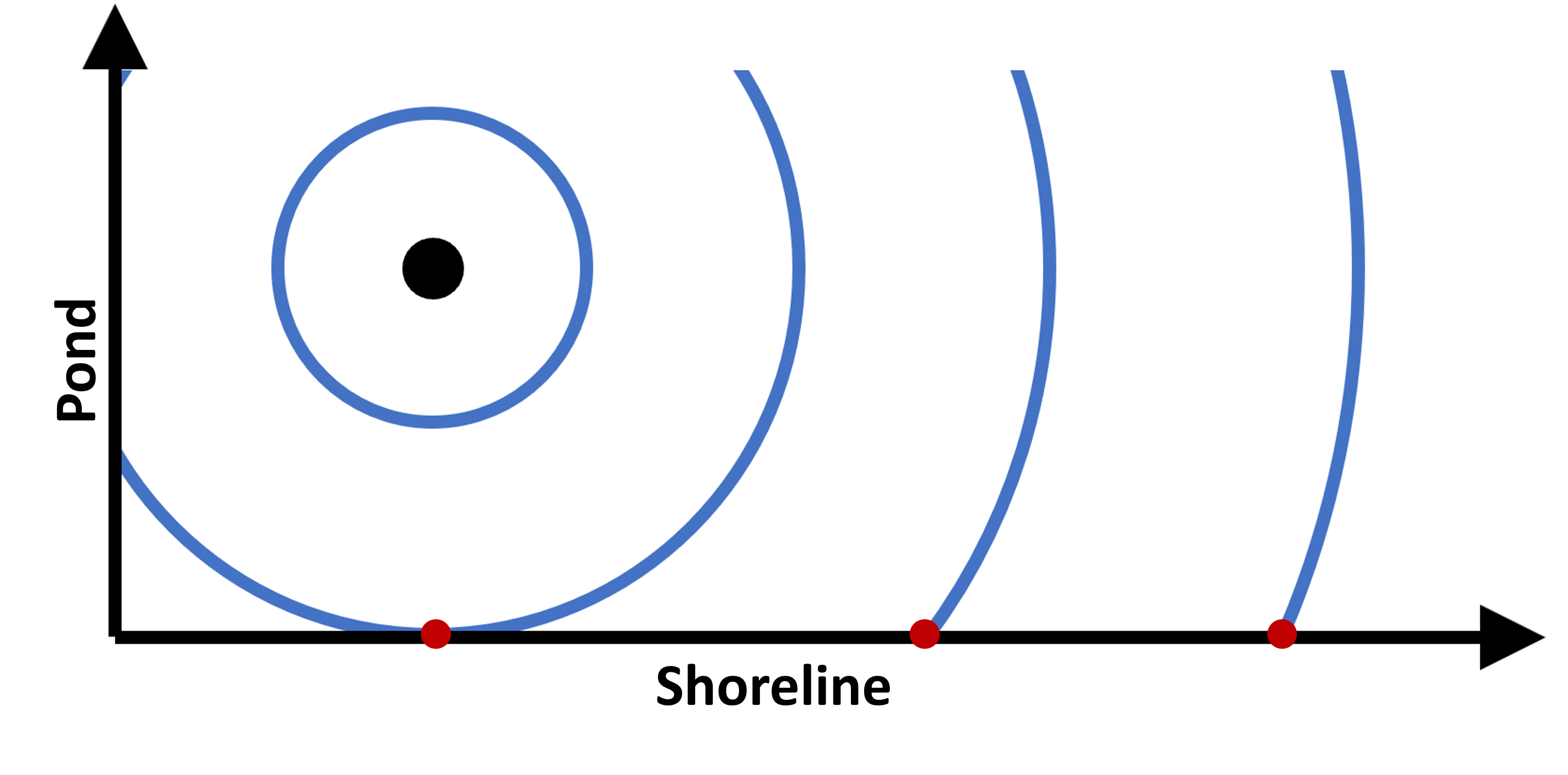}
    \end{center}
    \caption{Waves from the pebble in a pond. Each wave contacts the shoreline at a point (in red), and the point of contact moves down the shore to the right.}\label{fig:pebble}
\end{figure}

Suppose a pebble is thrown into an otherwise still pond. The splash from the pebble produces waves that form concentric circles, propagating outward from the location of the splash (see Figure~\ref{fig:pebble}). Eventually the waves reach the pond's shore. At any given instance,  each wave touches the shore at a single point, which appears to be running down the shore with (nearly) constant velocity. This creates the impression of an `object'  moving along the shore--whereas in fact  there is no `object' at all, only a moving location where the wave contacts the shoreline. 

Neither the original splash nor the emanating circular waves are apparent at the shoreline, which just sees a running point of contact. 

Is it possible that physical `objects' (like photons) are actually something like this? Is a photon in some sense a ``point of contact'' between spacetime and something that happens outside of spacetime? In the next section, we develop this idea a bit further by constructing a  simplified conceptual model. 

\subsection*{Introducing the `process dimension'}

One way to develop a more thorough understanding of a situation is to build a model. Models can be physical (like an architect's scale model of a building), or they can be visual or conceptual, like a diagram (which requires more imagination to appreciate). Model-building always involves making some simplifications, in order to reduce unnecessary complications and bring out essential details of the situation being modeled. 

Now we're trying to model a situation where events that occur outside of space and time ``collide'' with  our universe.
But it's difficult enough to visualize four-dimensional spacetime, let alone events that occur
somewhere outside!
To get a hold of this idea to begin with, we should try to simplify as much as possible, without excluding the essential details. In doing this, we may once again cite the example of Einstein, who is often quoted as saying ``Everything should be made as simple as possible, but no simpler.'' \footnote{The exact quote from Einstein is longer: ``...the supreme goal of all theory is to make the irreducible basic elements as simple and as few as possible without having to surrender the adequate representation of a single datum of experience.'' \cite{robinson_2018} } 

So let's consider a simplified ``universe'' in which there is only one space dimension and one time dimension.  Then we can represent  spacetime as a 2 dimensional plane, where the $x$ and $y$ axes represent space and time respectively. We now  introduce into the picture a third dimension, which we'll call the \emph{process dimension}, which is \emph{outside} of space and time. We suppose that events occur in this higher-dimensional ``process space'', and the influence of any event propagates down along the process dimension.

\begin{figure}[h]
    \centering
    \includegraphics[scale=0.45]{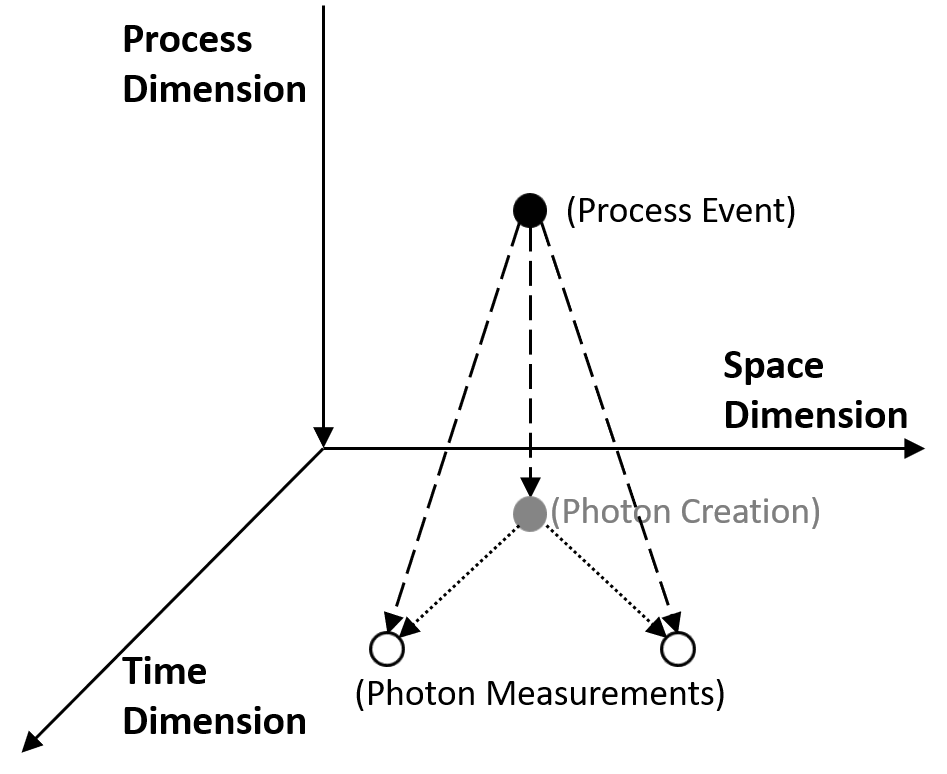}
    \caption{Representation of photon creation and measurement in the ``superspace'' picture. The process  (which occurs outside of spacetime) influences the photon creation and measurements: while in spacetime, we only see the creation and measurements. }\label{fig:superspaceEvent}
\end{figure}

Figure~\ref{fig:superspaceEvent} shows an event in  process space that occurs outside of spacetime, but nonetheless influences both the creation and subsequent measurements of two photons. For this perspective, it appears that nothing is actually ``travelling'' between the creation and detection of the photons: there's only the illusion of movement, which comes from the fact that both the creation and detection originate from a common source. 

This same picture can be generalized to describe any phenomenon that involves cause and effect. Suppose an event in our spacetime dimension occurs which seemingly ``causes'' another event. In our experience, the events happen sequentially. Yet, in the process dimension there is an unfolding process that  directly influences both the ``causing'' event and the `caused'' event, which produces the correlation between the two.

\section{Another view: the ``Many Worlds'' alternative}
 
 Another theory of reality is the `many-worlds interpretation' (MWI), which describes reality as having multiple alternative timelines that branch out at different events. A fanciful popular representation of this idea can be found in the ``Terminator'' movie series, in which time-traveling agents move along different timelines  to try and change past events (unfortunately, time travel is not actually a part of MWI!)  

MWI was first described as a possible explanation of quantum mechanics by Hugh Everett in 1957 in his Ph.D. thesis at Princeton University\cite{everett2015relative} (it is interesting however that Erwin Schr{\"o}dinger, one of the founders of quantum theory, had similar ideas previously\cite{gribbin}). Everett was  
trying to solve a major puzzle: it seemed that quantum particles like electrons and protons take the form of vague probability clouds, which instantaneously collapse suddenly every time a measurement is made. This so-called ``collapse of the wave packet'' is not part of the equations themselves, but is what evidently happens in our experience. In our quantum game show, this phenomenon is reflected in the fact that the choice of one door seems to instantaneously influence the result behind a different door, which is seemingly impossible.

Everett addressed this problem by suggesting that there is no ``collapse'' at all--instead, when a measurement is made all possible results do occur, but on different branches of reality. I only measure one result because the ``I'' that I'm aware of is following one of the branches. The different branches have different ``weights'', which explains why different outcomes appear to have different probabilities.

Despite its weirdness, the MWI has experienced a resurgence in popularity, and several prominent physicists are now  enthusiastic supporters. Some well-known examples include David Deutsch, Sean Carroll and Lev Vaidman, all of whom have  popularized the idea in articles\cite{vaidman2021Stanford}, books\cite{carroll2019something}, and/or videos (\cite{deutsch2009},\cite{Vaidman2014},\cite{Vaidman2020},\cite{Vaidman2021}\cite{deutsch2021}).

 \section{Which explanation is simpler?}
 
 We have presented two very different explanations for the quantum game show. Which explanation is right? Or are they both wrong?
 
 This may not be the right question to ask.  In fact, in physics there has never (so far) been an explanation that was ``right". For a long time, Newton's Three Laws were regarded as absolute truth. But around the beginning of the 20th century, it was found that these laws fail miserably outside of a very limited range of phenomena (which just so happened to be the phenomena that physicists had previously been able to study systematically). The very same thing has happened to every other famous equation in physics. We have no reason to believe that this situation will ever change--the weight of evidence points to the conclusion that we will never have a ``right'' equation in physics. 
 
 So instead of `Which is right?', perhaps we should ask other questions. Following Einstein's original quote in \cite{robinson_2018}, two questions we should be asking are: `Which agrees better with experience?' and `Which is simpler?' 
 Unfortunately, the question about experience is unhelpful in deciding between the process model and many worlds, because \emph{so far}, both predict the same thing. This situation may change however: David Deutsch claims there may be an experimental test of MWI \cite{gribbin}, and the process model should lead to tiny deviations from the quantum wavefunction \cite{thron2015accumulative}.

So for the present, we will focus on the question, ``Which explanation is simpler?''. Certainly such judgements are subjective, and individuals will have different predilections. 
A ``simpler'' theory should be easier to represent. So let us compare representations of the two theories. 
 
\begin{figure}[h]
    \centering
    \includegraphics[scale=0.45]{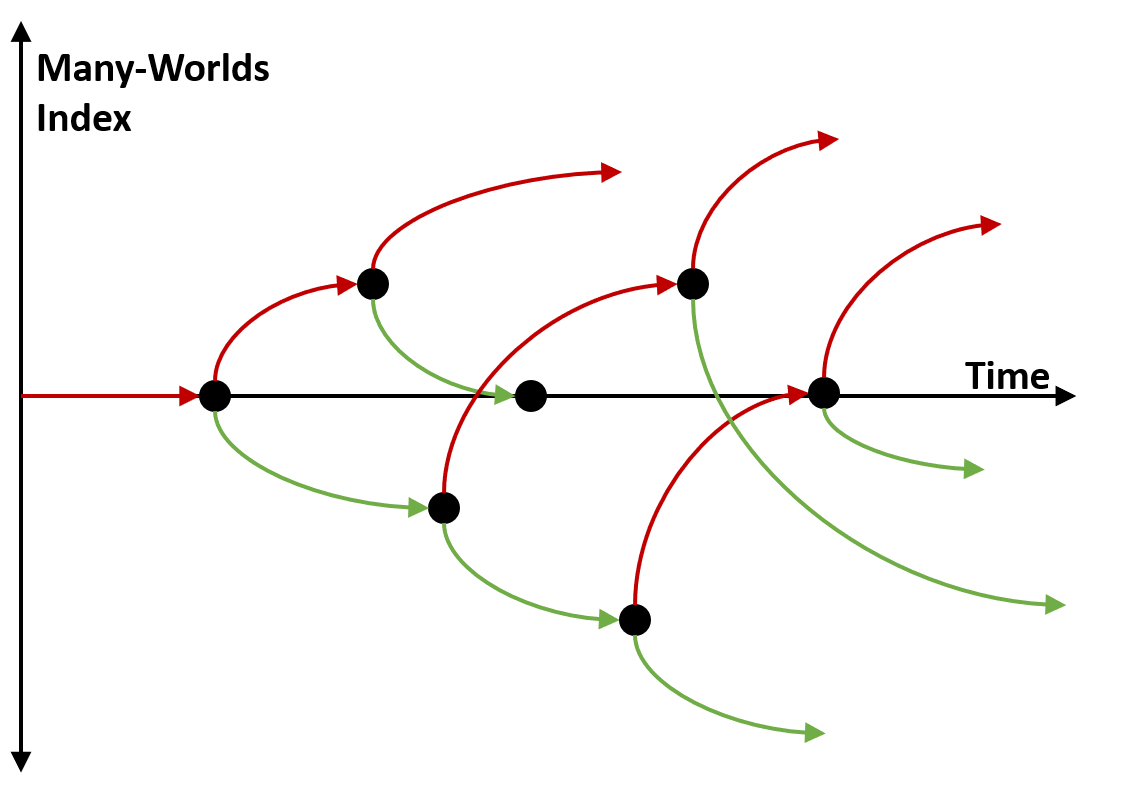}
    \caption{0-space + 1-time dimensional universe, according to the many worlds picture. }\label{fig:MW}
\end{figure}

Figure~\ref{fig:MW} depicts a  0-space + 1-time dimensional ``many worlds'' universe. The diagram is full of splitting timelines, where each split represents a spacetime event that has two possible outcomes (red or green). 
All of these timelines together constitute ``reality". The number of world lines increases exponentially as time progresses.  ``I" experience being on one of these branches, but alternate ``I"'s have different experiences on different branches. The events in ``my'' past and future are correlated because the current red/green state depends on previous branchings.

\begin{figure}[h]
    \centering
    \includegraphics[scale=0.45]{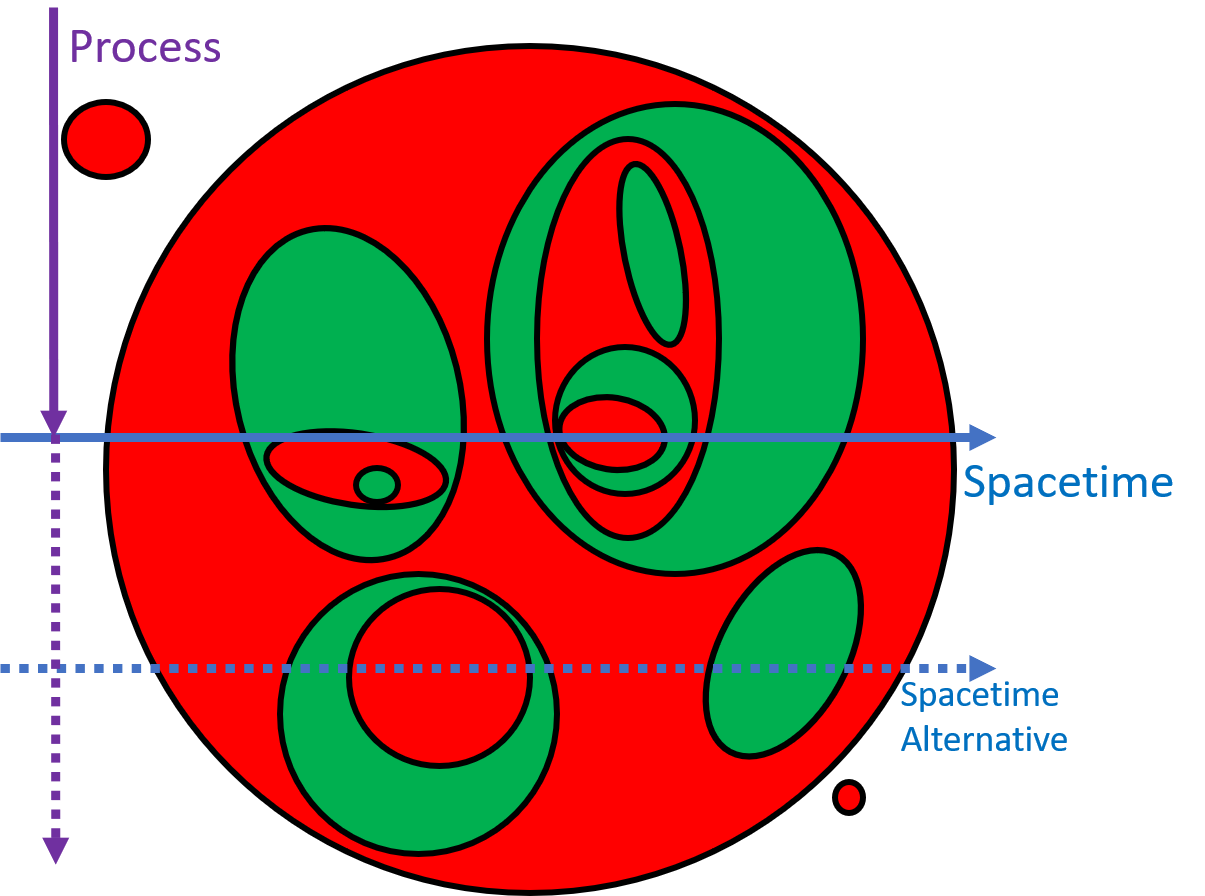}
    \caption{1-dimensional spacetime as a slice in 2-dimensional process space}\label{fig:slice}
\end{figure}

In contrast, Figure~\ref{fig:slice} 
represents the process space model for a 0+1 dimensional universe. Here we see `bubbles' in two dimensional process space, which represent the formation of cause-and-effect pairs. Instead of the wild branching of MWI, the physical spacetime universe appears as a single cross section taken from a larger picture. 

Which is simpler? Let the reader decide.

\section{Re-envisioning ``reality''}

Suppose that the process proposal is true, and the universe is a ``seashore'' for a larger ``ocean'' in which inaccessible processes take place.  What are the consequences of this new outlook? How does it affect how we look at the universe, at physics, and at ourselves?
The following subsection describes some of the conceptual consequences.

\subsection*{Ideas to consider}

\begin{itemize}

\item[\#1:]
\emph{Spacetime is a cross-section, not a process.}
We typically think of reality as a process that develops over time, and the sciences of physics, chemistry, biology etc. as describing processes in time. In virtually all equations in science that involve time, time is the independent variable and other quantities are functions of time (Einstein's equations in general relativity are the only exception that I can think of). 
Of course we do this because we experience the universe as an unfolding process in time.  But this is a terrible reason! We also experience the earth as stationary--and we all know where that led us. 

In the process space model, time is no longer singled out as special, and instead is placed on the same level as dimensions of space . Time seems different to us only because of the way we experience the universe.  For example, if you're riding a train things appear to be moving in one direction. But that's just because we're riding the train. If we were riding a different train, things would look like they're moving in a different direction. In the same way, things seems to develop in time just because of how we humans are `travelling' through spacetime.

From this angle, we may see that the complicated branching of MWI in Figure~\ref{fig:MW} can be traced back to the presumption that the universe is a process in time. If we attribute this process to another dimension, then the complication disappears.

Unfortunately, the process space model does not show us how to predict the long-term future or access the distant past. It can't help us build a time machine or a faster-than-light transporter. We're stuck in our universe, and the picture doesn't show us how to jump off into the process dimension to see the processes developing. But on the other hand, it doesn't say that we \emph{can't} jump off!

\item[\#2] \emph{There is no collapse of the wave packet.}
The conundrum posed by the so-called ``collapse of the wave packet'', was one of the primary motivations for MWI.  But in  the process theory, the conundrum disappears. The so-called wave packet is depicted as a statistical description of what goes on in the process dimension ``behind the scenes'', which sets the stage for what we see in spacetime. The process evolves in wavy fashion, but gives rise to sharp, specific physical outcomes in spacetime.   

\item[\#3:]
\emph{Entanglement is uncovered, not produced.}
Einstein's ``spooky action at a distance'' is a consequence of quantum entanglement, in which  particles or systems become mysteriously and invisibly ``linked'' due to  interactions. The process space model shows that interactions do not ``produce'' entanglement--rather, both the interaction and the measurements of entangled particles are vestiges of the same extra-universal process. 

\item[\#4:]
\emph{There are no causes without effects.} In the process space model, what we typically call ``causes'' and ``effects'' share a common origin outside of the universe. When the Hubble space telescope detects light from galaxies, the detection was not a ``afterthought'', but was ``predestined'' together with the light's emission thousands of years ago. This idea is actually not new--it has been proposed before, by Wheeler and Feynman\cite{wheeler1945interaction} and earlier by Tetrode\cite{tetrode1922causal}.  
However, they did not envision the cause-effect connection as the culmination of a process that takes place outside of the universe.

\item[\#5:]
\emph{There is no ``reality'' apart from observation. }
We typically think of an ``experiment'' as an observer looking at some phenomenon or physical system from the outside, and making impartial, passive  observations and measurements. This concept was already roughed up by quantum mechanics, which showed that quantum measurements are never passive, and always disturb the system they measure. The process space model goes one step further, and proposes that this is because there is no ``objective'' reality ``out there'' to measure.  The division between ``phenomenon'' and ``measurement'' is somewhat arbitrary--both are aspects of a larger whole within process space. 

It is interesting that there are several similar examples within physics of unifications between apparently disparate phenomena. There is for example the unification of electric and magnetic fields in classical electromagnetism; the  gauge field representations of matter and electromagnetism in quantum electrodynamics (see the wooden fish example in \cite{quodlibet2001}); and ``Grand Unified Theories'' and ``Theories of Everything'' in modern particle physics.)

One may raise the objection that reality must exist independent of observation, since different observers always observe the same reality. However, this is a magic trick that is due to the common origin of \emph{all} measurements within process space. 
 
\item[\#6]\emph{Spacetime is becalmed.}
Conventional quantum field theory takes the view that (quoting Lawrence Krauss in \cite{KraussNothing}) ``Empty space is a boiling, bubbling brew of virtual particles that pop in and out of existence in a time scale so short that you can't even measure them.'' 
The process space model interprets this supposed storm of activity  as a mathematical trace of the unfolding process located outside of accessible spacetime.

\item[\#7:] \emph{Free will returns!}
There is a pervasive ``scientific'' idea (championed by Sam Harris \cite{harris2012free}, among others) that past physical processes dictates our present and future. This is an illusion, says the process space model. You are not determined by your past, nor merely the product of circumstances. Both you and previous circumstances come from somewhere else. 
There are hidden processes that gave rise to make you who you are in this world, but there are not physical processes in this universe. 

One might suggest that the process theory just displaces determinism into another dimension, and that we are still the outcome of processes outside of ourselves. Not so fast. In the process space model, the meaning of ``self'' is no longer so clear. How far do ``I'' extend into the process dimension? To say that ``I'' am confined to spacetime is like saying that an iceberg is only the visible part that sticks above the water.

\item[\#8:] \emph{The quest for an ultimate ``Theory of Everything'' is a rainbow chase.}
Consider the evolution of the conception of the physical world in Western intellectual history.
Greek philosophers conceived of solid objects as space-filling entities having extension and rigidity as fundamental properties. Much later, objects were re-envisioned as consisting of ensembles of jiggling atoms, held together by electromagnetic forces. Later, atoms themselves became  mostly empty space, consisting of much smaller particles (with electrons having no size at all). Then these particles turned into quanta, which are a strange amalgam of particle and wave. Later, quanta turned into resonances of a field, and different particles corresponded to different components of a multdimensional field.  Then we had vibrating strings, then membranes. 
This process is clearly not going to end. 
 
 By now, it should be abundantly clear that every physical theory is simply convenient conceptualization of our current level of interaction with the universe. We are never getting any closer to ``true reality'', which is a concept that should be banished from the scientific/philosophical vocabulary.  The best we can do is provide a series of analogies, supported by mathematical substructures. These are all temporary structures, and  will always be superseded. In a related context,  Ludwig Wittgenstein gave the analogy of a ladder, which is used to climb up to a higher level and is then discarded as unnecessary.
 
 \item[\#9:] \emph{Science explains almost nothing.}
Physical science is concerned with relationships between events, phenomena, and conditions in spacetime.  But the process space model shows that spacetime itself is only a single slice of process space, and, as such, has no direct access to the bulk of processes that rule the universe. As such, science can only offer explanations and predictions that are highly generalized or tightly localized in space and time. Consider for example the famous ``baby picture of the universe'' (see \cite{NASA}), which shows variations in the temperature of the microwave background radiation of the universe on the order of 200 microkelvins. It is indeed a great triumph that the cosmic inflation theory predicts  the observed blotchiness. However, it cannot even hope to explain the location or extent of a single blotch. At the small end of the scale, lattice quantum chromodynamics (QCD) simulations of a single neutron require computing power equivalent to 6,000 laptops operating continuously for a century \cite{berkeley2018}.

\item[\#10:]\emph{Consciousness may be extra-physical.}
 There are phenomena within the physical world that do not have their origin within the physical world, and thus cannot be constructed by  physical procedures. Consciousness may be one such phenomenon (and angels?). So we may not be able to build conscious computers.

\end{itemize}

\subsection*{More in depth}
Mathematical and theoretical aspects of the process theory for ordinary quantum mechanics (not quantum field theory) has been developed in a series of papers (\cite{thron2013signal},\cite{thron2015b}). The second paper also explains physical consequences of the theory that differ from conventional quantum mechanics, which could conceivably be used to test the theory.

A video series that presents the theory on a more popular level has been placed on YouTube \cite{Chrisyoutube}. Note however this series does contain some inaccuracies.

\bibliographystyle{plain}
\bibliography{main}

\begin{thebibliography}{10}

\bibitem{aspect1982experimental}
Alain Aspect, Jean Dalibard, and G{\'e}rard Roger.
\newblock Experimental test of bell's inequalities using time-varying
  analyzers.
\newblock {\em Physical review letters}, 49(25):1804, 1982.

\bibitem{carroll2019something}
Sean Carroll.
\newblock {\em Something deeply hidden: quantum worlds and the emergence of
  spacetime}.
\newblock Dutton, 2019.

\bibitem{doyle2014}
Arthur Conan~Doyle.
\newblock {\em The Sign of the Four}.
\newblock Penguin Classics, London, 2014.

\bibitem{deutsch2009}
David Deutsch.
\newblock Apart from universes.
\newblock Vimeo, 2009 [Online].
\newblock \url{https://vimeo.com/5490979}.

\bibitem{deutsch2021}
David Deutsch.
\newblock Multiple worlds and our place in them.
\newblock Youtube, 2021 [Online].
\newblock \url{https://www.youtube.com/watch?v=b_6vYwCkIpc}.

\bibitem{everett2015relative}
Hugh Everett.
\newblock ``relative state'' formulation of quantum mechanics.
\newblock In B.~DeWitt and N.~Graham, editors, {\em The Many Worlds
  Interpretation of Quantum Mechanics}, pages 141--150. Princeton University
  Press, 2015.

\bibitem{NASA}
Megan Gannon.
\newblock New `baby picture' of universe unveiled.
\newblock Space.com, 2012 [Online].
\newblock \url{https://www.space.com/19027-universe-baby-picture-wmap.html}.

\bibitem{gribbin}
John Gribben.
\newblock The many-worlds theory, explained.
\newblock MIT Press, 2020 [Online].
\newblock \url{https://thereader.mitpress.mit.edu/the-many-worlds-theory/}.

\bibitem{harris2012free}
Sam Harris.
\newblock {\em Free will}.
\newblock Simon and Schuster, 2012.

\bibitem{berkeley2018}
Glenn~Roberts Jr.
\newblock Supercomputers provide new window into the life and death of a
  neutron.
\newblock Berkeley Lab, 2018 [Online].
\newblock
  \url{https://newscenter.lbl.gov/2018/05/30/supercomputer-simulations-new-window-lifetime-death-neutron/}.

\bibitem{KraussNothing}
Lawrence Krauss.
\newblock Lawrence krauss on `a universe from nothing'.
\newblock National Public Radio, 2012 [Online].
\newblock
  \url{https://www.npr.org/2012/01/13/145175263/lawrence-krauss-on-a-universe-from-nothing}.

\bibitem{maccone2013simple}
Lorenzo Maccone.
\newblock A simple proof of bell's inequality.
\newblock {\em American Journal of Physics}, 81(11):854--859, 2013.

\bibitem{Aspectbrief}
Frank Rioux.
\newblock A brief description of aspect\'s experiment.
\newblock LibreTexts, 2020.
\newblock
  \url{https://chem.libretexts.org/Bookshelves/Physical_and_Theoretical_Chemistry_Textbook_Maps/Supplemental_Modules_(Physical_and_Theoretical_Chemistry)/Quantum_Tutorials_(Rioux)/Quantum_Teleportation/354\%3A_A_Brief_Description_of_Aspect's_Experiment}.

\bibitem{robinson_2018}
Andrew Robinson.
\newblock Did einstein really say that?
\newblock {\em Nature}, 557(30), 2018 [Online]. doi:
  https://doi.org/10.1038/d41586-018-05004-4.

\bibitem{tetrode1922causal}
Hugo~Martin Tetrode.
\newblock On the causal connection of the world, an extension of classical
  dynamics.
\newblock {\em Zeitschrift fur Physik}, 10:317--328, 1922.

\bibitem{quodlibet2001}
Chris Thron.
\newblock God is light, sin is entropy: Physical analogies for biblical
  concepts.
\newblock Fifth Interdisciplinary Conference on Science, Technology and
  Religious Ideas, Kentucky State University, 1994 [Online].
\newblock \url{https://crosspollen.neocities.org/sin_entropy.htm}.

\bibitem{Chrisyoutube}
Chris Thron.
\newblock Is cause and effect an illusion? arguments from physics.
\newblock Youtube, 2015 [Online].
\newblock
  \url{https://www.youtube.com/playlist?list=PL2uooHqQ6T7PgUzsM95xUWfBuyeGBUXzx}.

\bibitem{thron2013signal}
Chris Thron and Johnny Watts.
\newblock A signal-processing interpretation of quantum mechanics.
\newblock {\em The African Review of Physics}, 8, 2013.

\bibitem{thron2015accumulative}
Christopher Thron.
\newblock An accumulative model for quantum theories.
\newblock {\em arXiv preprint arXiv:1507.03944}, 2015.

\bibitem{thron2015b}
Christopher Thron.
\newblock An accumulative model for quantum theories.
\newblock {\em Electronic Journal of Theoretical Physics}, 12(33), 2015.
\newblock \url{http://www.ejtp.com/ejtpv12i33.html} (see also
  \url{https://arxiv.org/abs/1507.03944}.

\bibitem{Vaidman2014}
Lev Vaidman.
\newblock From bell inequalities to many worlds interpretation.
\newblock YouTube, 2014 [Online].
\newblock \url{https://www.youtube.com/watch?v=jKGuGptafvo}.

\bibitem{Vaidman2020}
Lev Vaidman.
\newblock In favor of the many worlds interpretation.
\newblock CRNS-Grenoble, 2020 [Online].
\newblock \url{https://www.youtube.com/watch?v=CE3rzGHUjqI}.

\bibitem{Vaidman2021}
Lev Vaidman.
\newblock In the many-worlds interpretation.
\newblock Harvard University, 2021 [Online].
\newblock \url{https://www.youtube.com/watch?v=1Yx-wQYODiA}.

\bibitem{vaidman2021Stanford}
Lev Vaidman.
\newblock Many-worlds interpretation of quantum mechanics.
\newblock Stanford Encyclopedia of Philosophy, 2021 [Online].
\newblock \url{https://plato.stanford.edu/entries/qm-manyworlds/}.

\bibitem{wheeler1945interaction}
John~Archibald Wheeler and Richard~Phillips Feynman.
\newblock Interaction with the absorber as the mechanism of radiation.
\newblock {\em Reviews of modern physics}, 17(2-3):157, 1945.

\end{thebibliography}

\end{document}